\documentclass[12pt]{amsart}
\usepackage[margin=1in]{geometry}

\usepackage{amsmath,amsfonts,amssymb,amsthm}
\usepackage{graphicx}
\usepackage[dvipsnames]{xcolor}
\usepackage{wrapfig}
\usepackage{url}
\usepackage{caption}
\usepackage{subcaption}
\usepackage[justification=centering]{caption}
\usepackage{bbold}

\usepackage{hyperref}\hypersetup{colorlinks=True, urlcolor=cyan}
\usepackage{algorithm} 
\usepackage{algpseudocode}

\title[Opinion Dynamics Related To The Covid-19 Pandemic]{Opinion Dynamics Related To Covid-19 vaccine hesitancy and mega-influencers}
\date{8 April 2022}
\author{Anna Haensch}
\address[Anna Haensch]{Tufts University, Data Intensive Studies Center}
\email[Corresponding author]{anna.haensch@tufts.edu}

\author{Natasa Dragovic}
\address[Natasa Dragovic]{Tufts University, Department of Mathematics}
\email{natasa.dragovic@tufts.edu}

\author{Christoph B\"orgers}
\address[Christoph B\"orgers]{Tufts University, Department of Mathematics}
\email{christoph.borgers@tufts.edu}

\author{Bruce Boghosian}
\address[Bruce Boghosian]{Tufts University, Department of Mathematics}
\email{bruce.boghosian@tufts.edu}

\begin{document}

\maketitle

\begin{abstract}
 Covid-19 vaccines are widely available in the United States, yet our Covid-19 vaccination rates have remained far below $100\%$. Data from the CDC show that even in places where vaccine acceptance was proportionally high at the outset of the Covid-19 vaccination effort, that willingness has not necessarily translated into high rates of vaccination over the subsequent months. We model how such a shift could have arisen, using parameters in agreement with data from the state of Alabama. The simulations suggest that in Alabama, local interactions would have favored the emergence of tight consensus around the initial majority view, which was to accept the Covid-19 vaccine. Yet this is not what happened. We therefore add to our model the impact of mega-influencers such as mass media, the governor of the state, etc. Our simulations show  that a single vaccine-hesitant mega-influencer, reaching a large fraction of the population, can indeed cause the consensus to shift radically, from  acceptance to  hesitancy. Surprisingly this is true even when the mega-influencer only reaches individuals who are already somewhat inclined to agree with them, and under the conservative assumption that individuals give no more weight to the mega-influencer than they would give to a single one of their friends or neighbors. Our simulations also suggest that a competing mega-influencer with the opposite view {\em can} shift the mean population opinion back, but under some conditions {\em cannot} restore the tightness of consensus 
 around that view. Our code and data are distributed in the ODyN (Opinion Dynamic Networks) library available at \url{https://github.com/annahaensch/ODyN}.
\end{abstract}

\section{Introduction}

Opinions drive human behavior \cite{review_paper}, and opinion formation is a complex multi-scale process, involving characteristics of the individual, local interaction of individuals, social media, mass media etc.  Opinion dynamics have been modeled using approaches inspired by physics \cite{hofstad_networks_book}.  A survey of the 
literature on opinion dynamics can be found in \cite{survey_general}. In this study, we use the 
example of 
opinions about Covid-19 vaccination, in part because the topic is of urgent current interest, but
also because data are plentiful (see for instance \cite{cdc_hesitancy_trends}, which
we use as a source in our simulations), and a significant shift in opinions appears to have occurred
in the United States in a short time span. We focus on two scales, local interactions (conversations with friends, family, neighbors, colleagues) and global 
influencers such as mass media and prominent politicians, whom we refer to as 
{\em mega-influencers}.  Social media can belong to either category. A discussion of the merits or perils of Covid-19 vaccination among 25 Facebook friends could be viewed as a local
interaction, whereas a Twitter account owner with millions of followers is a mega-influencer. In the model of this paper, social media will not appear 
explicitly.  Despite the apparent geography-less nature of the online world, studies have shown \cite{lengyel_2015} that geographic distance is still a key component in the formation and maintenance of social networks.  Therefore, in the present paper we situate individuals in physical and opinion distance, and take this as a model for our network.

Covid-19 vaccination rates over time  have been closely tracked by the US Centers for Disease Control and Prevention at both the national and county levels \cite{cdc_hesitancy_trends,cdc_county_trends, cdc_national_trends,cdc_data}.
By April 19, 2021, vaccination was approved for everyone in the US age 16 and older. Despite the fact that the vaccine was free to all residents of the US, many factors impeded widespread vaccination.  There was a lack of availability and access to vaccines in rural areas \cite{murthy_rural}, social and economic factors impacted vaccine hesitancy \cite{simas_overcoming}, and there were also targeted misinformation campaigns and politicization of issues surrounding the vaccine \cite{rabb_plosone}.  Fear of adverse health effects related to the Covid vaccine might also be factor, despite the fact that studies have shown that adverse effects are quite rare \cite{beatty_2021}. For purposes of this study, we intend to focus entirely on the social dynamics of vaccine hesitancy as one possible explanation for vaccination levels lagging below what is expected.

\begin{figure}
    \centering
    \includegraphics[width = .9\linewidth]{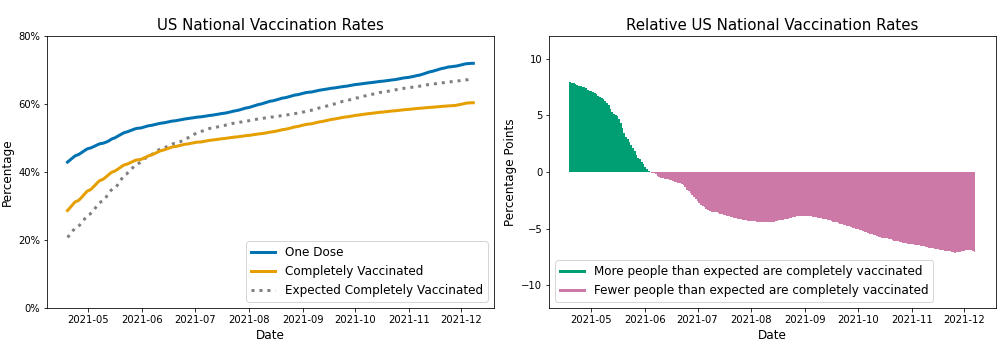}
    \caption{
    Left: U.S. national vaccination rates. ``Completely vaccinated" means one dose of Johnson \& Johnson or two doses of Pfizer/BioNTech or Moderna. ``Expected completely vaccinated" curve obtained by shifting  ``one dose" curve forward by 42 days. Right: Actual minus expected vaccination rates.}\label{fig:us_trends}
\end{figure}

\begin{figure}
    \centering
    \includegraphics[width = .9\linewidth]{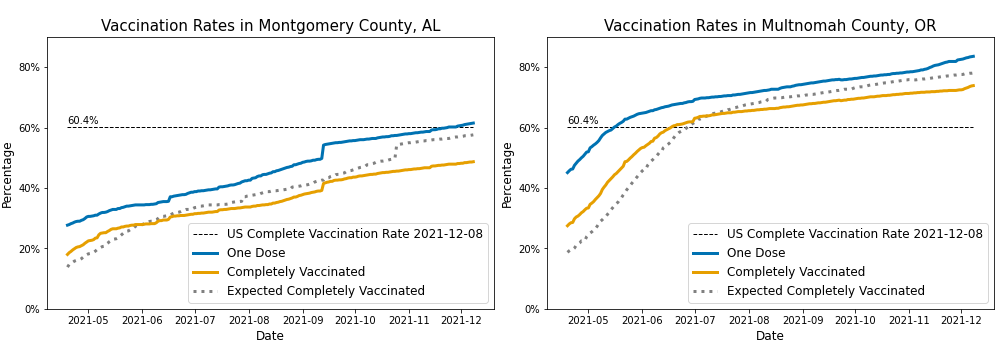}
    \caption{
    The vaccination rate is below the national average in Montgomery County, AL (left),
    above it  in Multnomah County, OR (right). 
    Yet in both counties, a shift towards greater hesitancy appears to have 
    occurred, indicated by the fact that the orange curves have fallen below
    the dotted grey ones.
    }\label{fig:county_trends}
\end{figure}

Fig~\ref{fig:us_trends} shows the proportion of {\em partially} vaccinated people in the US as a function of time, beginning on April 19, 2021 (blue),
and also the proportion of {\em fully} vaccinated people (orange).  The proportion of fully
vaccinated people around December of 2021 was around $60\%$.  According to the terminology introduced by the CDC and widely adopted throughout Covid related news coverate, a person is considered  {\em fully vaccinated} if they have had either one shot of the Johnson \& Johnson vaccine or two shots of the Pfizer/BioNTech or Moderna vaccine, where the second shot must follow within 42 days of the first \cite{kriss_completion}.  Fig~\ref{fig:us_trends} also shows that the actual number of people who achieved complete vaccination at some point began to fall short of the expected number.  In other words, despite having access to the vaccine, and receiving their first dose, people did not always follow through with their second dose (orange in Fig~\ref{fig:us_trends}). There appears to have been a shift towards greater vaccine hesitancy over time. Similar trends can be seen at the county level, in many counties throughout the
United States (Fig~\ref{fig:county_trends}).
In this paper we examine how such a shift might have come about.

Opinions about issues such as Covid-19 vaccination are formed in part
by people talking 
to their families,  friends,  colleagues,  etc. This is the sort of mechanism that
for instance the Hegselmann-Krause (bounded confidence) model
\cite{hegsel_krause_2002} aims to capture. In \cite{hegsel_krause_2015}, Hegeselmann and Krause modify their model further to consider the impact of radicals on opinions. In their model, radicals are individuals (or groups of individuals) that hold extreme and persistent opinions on a single end of the opinion spectrum.  By contrast, our model aims to capture geospatial as well as opinion space dynamics in the presence of what we've called {\em mega-influencers} on both ends of the spectrum.  We endeavor to explore how {\em mega-influencers} such as mass media and prominent politicians interact
with Hegselmann-Krause-type dynamics.

We define a random graph in which the vertices represent individuals, and
{\em directed} edges indicate who influences whom. (When individual $v$'s opinion influences that of individual $u$, this does not necessarily imply that $u$ also 
influences $v$.) In our model,  different people
have different levels of influence, as in a Chung-Lu random graph
 \cite{chung_lu1,chung_lu2,chung_lu3}. Further, in our model people are more likely to be connected to those  who are spatially closer to them, as in the geometric inhomogeneous random
 graphs of Brigmann {\em et al.}
  \cite{girgs_algorithm_paper}.  As in work by Mathias et al. \cite{mathias_2016} we include the influence of a small number of individuals holding extreme, immutable beliefs, we call them {\em mega-influencers} (related literature also uses the term {\em radicals}).  Our treatment differs from \cite{mathias_2016} in that the underlying network is spatial and mimics a real word network (see Section \ref{methodology} for a more robust discussion on the network parameters). For a review of spatial networks, we direct the reader to Barth{\'e}lemy's comprehensive survey on the topic \cite{barthelemy_2011}.  The networks that we propose in the present paper are similar to the hidden variable model for spatial networks presented in \cite[Section 3]{barthelemy_2011}, but futher enhanced to include bounded confidence.

To initialize the network model every person starts with their own {\em belief}, defined as their position in opinion space, and individual
beliefs 
evolve according to the Hegselmann-Krause model  \cite{hegsel_krause_2002},
which  assumes that 
 people are influenced only by opinions similar to theirs.  
 The Hegselmann-Krause model has been studied extensively in the literature \cite{literature_hk1,literature_hk2, literature_hk3, literature_hk4}. It is just one of several popular opinion dynamics models; see \cite{other_model1, other_model2, other_model3, gargiulo_2016} for other examples. 

In the simulations presented here, the spatial domain is a triangle, 
and the spatial locations of individuals are independent of each other and uniformly
distributed. The code in the ODyN (Opinion Dynamic Networks) library (\url{https://github.com/annahaensch/ODyN}) 
also allows simulations on unions of triangles, where the number of individuals
in each triangle is chosen to be random, Poisson-distributed, with an expected value proportional
to the area of the triangle, possibly with different constants of proportionality for
different triangles. In short, the spatial locations in the ODyN library are
a Poisson point process with a possibly space-dependent rate. 
We plan to use this code to study irregular
shapes such as US states by (approximate) triangulation in the future; see 
 Fig~\ref{fig:triangulation_figure}.

Simulations of our model suggest that in Alabama, local interactions would have favored consensus around the initial majority view, which was to accept the Covid vaccine. We show that the addition of a vaccine-hesitant mega-influencer can cause the consensus to shift radically, from acceptance to hesitancy. A competing mega-influencer with the opposite view can bring the mean of the opinion distribution back, but the spread of opinions
will be significantly greater with both mega-influencers present than without either
of them. Our code and data are distributed in the ODyN library.

\section{Model}\label{model}

\subsection{Directed graph encoding who influences whom} Let $N$ be a positive integer, and consider $N$ individuals. We use letters such as $u$ and $v$ (for ``vertex"), $1 \leq u, v \leq N$, 
to label individuals. We will construct a random {\em directed} graph in which the individuals are the vertices, with an arrow (a directed edge) from individual $v$ to individual $u$
indicating that $v$ influences the opinion of $u$. We write 
$$
p_{uv} = \mbox{probability of an arrow from $v$ to $u$}.
$$
We do not assume symmetry: $p_{vu}$ need not be equal to $p_{uv}$.

\subsection{Spatial locations}
\begin{figure}
    \centering
    \includegraphics[width=.75\linewidth]{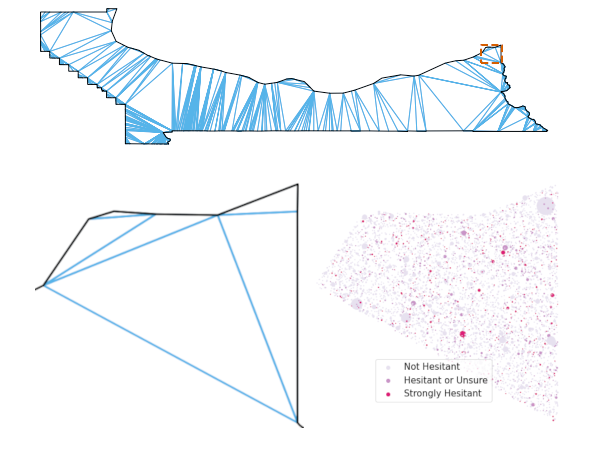}
\caption{The polygonal boundary of Multnomah County, Oregon (top) is triangulated prior to populating with a Poisson point process.  Triangular regions (bottom left) are populated with the proper population density and belief proportion (bottom right).  Triangles are generated to connect all vertices on the polygonal region that represents the county geometry.  In some cases boundary points appear colinear at this resolution, although in reality they are not. }\label{fig:triangulation_figure}
\end{figure}

This part of our model is inspired by  \cite{girgs_algorithm_paper}, although several of the details are
different here. 
We assign to individual $v$ a random spatial location $X_v$ in a polygonal domain $D$ in the plane.  In the code available through ODyN, 
$D$ is assumed to be a union of triangles, and the number of individuals per triangle
is taken to be random with Poisson distribution, with a rate that can be different
for different triangles. The locations of individuals within 
each triangle are then assumed to be independent and random with uniform distribution (see Fig~\ref{fig:triangulation_figure} for an example). We use triangles because
they are a flexible way of approximating more complicated shapes, and it is straightforward to generate uniformly distributed random points in a triangle.

In the simulations presented here, we simply take
$D$ to be a single triangle, fix $N$, and let the locations $X_1,X_2,\ldots,X_N$
of the individuals be independent, uniformly distributed points in $D$.
We assume
that $p_{uv}$ is a decreasing function of the euclidean distance $\| X_u - X_v \|$.

\subsection{Influence weights.} \label{sec:weights}

Following Chung and Lu \cite{chung_lu1}, we assign a random {\em influence weight} $W_v > 0$ to each $v$. This weight
determines how likely others are to listen to $v$, not how much weight they assign to $v$'s opinion; the probability $p_{uv}$ is 
an increasing function of $W_v$. 

We assume $W_v$ to be a heavy-tailed random variable that is always
greater than 1. Specifically, we assume that for any 
$x>1$, 
\begin{equation}
\label{eq:complementary_distribution}
P(W_v>x) = \frac{1}{x^{\gamma}},
\end{equation}
with some exponent $\gamma>0$. (The parameter $\beta$ of \cite{chung_lu1} is $\gamma+1$.) To generate a random number $W_v$ with the complementary 
distribution function (\ref{eq:complementary_distribution}), draw a uniformly distributed random number $U \in (0,1)$, then 
set 
$$
W_v = U^{-\frac{1}{\gamma}}. 
$$
We will choose a value of $\gamma$ that makes the mean of the distribution of the 
$W_v$ finite: $\gamma>1$. Given this constraint, however, we will choose $\gamma$ to
make the variance of the distribution infinite, so that 
outlying values of $W_v$ become fairly common. The variance is infinite for 
$1 < \gamma \leq 2$, and since within this range, we don't expect the precise value of $\gamma$ to have a qualitative impact on our results, we choose $\gamma = 1.5$.

\subsection{Opinion scores.}\label{sec:opinions} We assign to each individual $v$ an 
{\em opinion score} $H_v$ between $0$ and
$2$, reflecting 
their view on Covid-19 vaccines, ranging from $H_v=0$ (no hesitancy) to $H_v=2$ (strong hesitancy). 
We assume that $p_{uv}=0$ if $|H_u - H_v| \geq b$, where $b>0$ is a threshold. That is, we assume
that $v$ cannot have any impact on $u$'s opinion  if $u$ and $v$ have starkly different views.

\subsection{Overall formula for the connection probabilities.} We define
\begin{eqnarray}\label{eqn:prob_influence}
p_{uv} = \min \left( 1,  \frac{1}{\left( 1 + \| X_u - X_v \|/\lambda \right)^\delta} ~ W_v^\alpha ~ \mathbb{1}_{|H_u-H_v|<b}  \right)
\end{eqnarray}
where $\mathbb{1}$ denotes the indicator function. The parameter $\lambda>0$ is a reference
length; we take it to be the diameter of the spatial domain. The parameters
$\alpha>0$ and $\delta>0$ determine the importance of influence weight and spatial
proximity, respectively.

\subsection{Initialization} The influence weights $W_u$ and spatial locations $X_u$ are independent random numbers, 
chosen as outlined above. The opinion scores $H_u$ change with time; see Section \ref{sec:hegselmann_krause}. We assign a random initial opinon 
score  to each individual $u$. These assignments 
are made independently of each other, and independently of the $H_u$ and $W_u$. Initial opinion scores 
take on integer values only: 0, 1, or 2. The probability that an individual $u$ is assigned the initial opinion score $k$ 
equals $p_k$, $k=0,1,2$, where the 
 $p_k$ are chosen to reflect  publicly available data. At later times, we allow the $H_u$ to 
take on any values in the interval $[0,2]$.

\subsection{Hegselmann-Krause dynamics}
\label{sec:hegselmann_krause}
Denote the opinion scores after $t$ time steps by $H_u(t)$. (We take $t$ to be a non-negative integer here.) Then $H_u(t)$ is the average
of those $H_v(t-1)$ for which either $v=u$, or there is an arrow pointing from $v$ to $u$ at time $t-1$. In words, $u$ 
averages their own opinion with the opinions of those whom $u$ is influenced by.
This is  the Hegselmann-Krause model \cite{hegsel_krause_2002}.  

Since the probabilities $p_{uv}$ depend on $H_u - H_v$, 
they, too, are time-dependent. The connections in the random graph are  re-drawn after each time step. The motivation for re-drawing the connections after each timestep is to simulate the fact that people don't necessarily speak and interact with the same people every day.

\subsection{In-degree and clustering coefficient.} The in-degree of an individual $u$ is the number of individuals $v$ 
who influence $u$, that is, the number of $v$ for which there is an arrow from $v$ to $u$. We will keep track of the average in-degree.  As the graph is time-dependent, so is the average in-degree. Since every outgoing arrow for one vertex is an incoming arrow for another vertex, the average in-degree is the same as the average out-degree.

The {\em clustering coefficient} of an individual $u$ is defined as follows. Denote by $k$ the number of 
individuals who influence $u$. 
If $k \leq 1$, then $u$ has clustering coefficient $0$. Otherwise, determine for each of  the $k(k-1)$ ordered pairs of individuals who  influence $u$ 
whether it is connected, that is, whether there is an arrow from the first to the second. The fraction of connected pairs is the clustering
coefficient of $u$.  

\begin{figure}
\centering
  \includegraphics[width=5in]{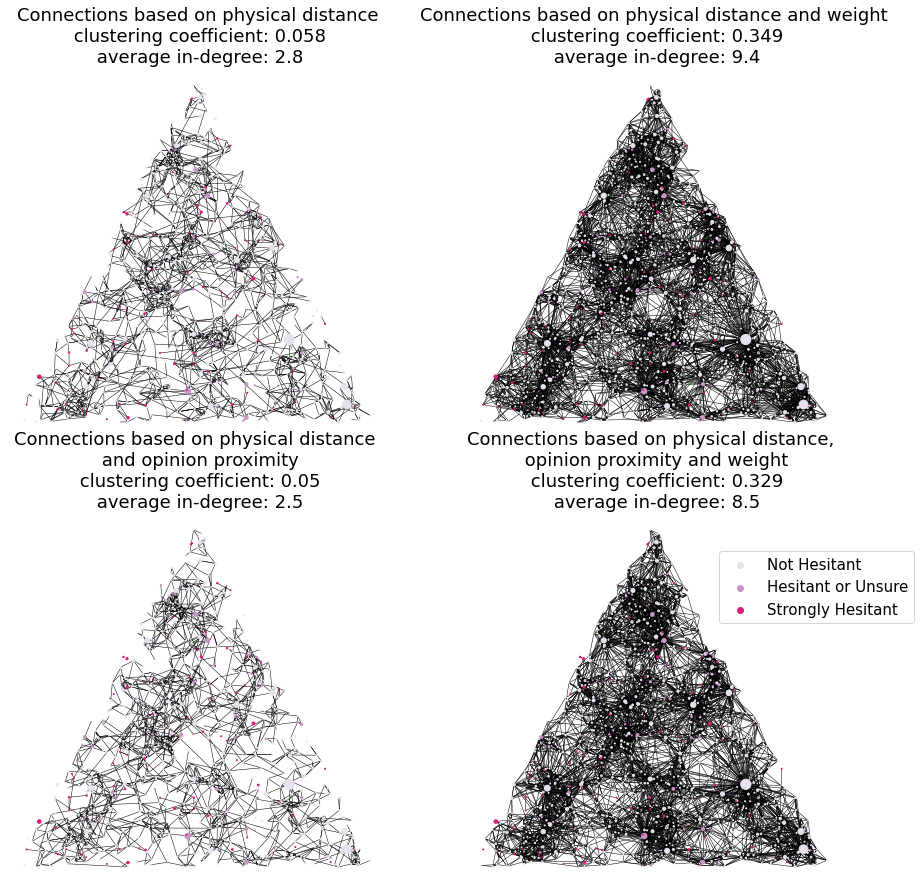}
  \caption{Four different initialized network model with $n=1000$ individuals, symmetric initial beliefs and varying active components informing the probability of adjacency. The initialization on the lower right is the one we actually use in our experiments.}
  \label{fig:ex_neighbor_changing}
\end{figure}

Both the average in-degree and the average clustering coefficient provide a way of evaluating whether our graphs are 
realistic, and therefore help in setting parameters.
The average in-degree should not be unrealistically high or low, and the average clustering coefficient should not be 
too low.  To see the overall effect of the inclusion of weights and beliefs on the clustering coefficients and in-degree, we have initialized several populations varying components informing the probability of adjacency (i.e. opinions, weights, or both), see Fig~\ref{fig:ex_neighbor_changing}. The network model we work with is on the bottom right of  Fig~\ref{fig:ex_neighbor_changing}. 

\subsection{Mega-influencers.} We add to the model two {\em mega-influencers}, one with opinion score $0$, referred to as 
the {\em left mega-influencer}, and the other
with opinion score $2$,  the {\em right mega-influencer}. One might think of these as modeling mass media outlets, outspoken governors, etc.

The impact of the mega-influencers is modeled as follows. To each individual $u$, we assign two random numbers 
$L_u$ and $R_u$, with
$$
L_u  = \left\{ \begin{array}{cl} 1 & \mbox{with probability $p_L$}, \\
0 & \mbox{otherwise}, 
\end{array}
\right. \hskip 30pt  \mbox{and} \hskip 30pt
R_u  = \left\{ \begin{array}{cl} 1 & \mbox{with probability $p_R$}, \\
0 & \mbox{otherwise}, 
\end{array}
\right. 
$$
where $p_L$ and $p_R$ are further model parameters, with $0 \leq p_L, p_R \leq 1$.
 If $L_u=1$, then $u$ is susceptible to the left mega-influencer. In that case, while $H_u \in [0,\epsilon)$, where $\epsilon>0$ is another mode
 parameter, the opinion score of the left mega-influencer, namely $0$, 
will be added to the opinions over which $u$ averages in each step of the Hegselmann-Krause dynamics.  Similarly, 
if $R_u=1$, then $u$ is susceptible to the right mega-influencer. In that case, 
while $H_u \in (2-\epsilon, 2]$, the opinion score of the right mega-influencer, namely $2$, will be added to the opinions over which 
$u$ averages in each step.
The parameters $b$ and $\epsilon$ play similar roles, for local interactions 
and for mega-influencers, respectively. In the code, they need not be the same, 
but in the simulations presented here, they were the same.

Note that our model assumes that to $u$, mega-influencers do not carry more weight than friends or neighbors. 
The very considerable effect of mega-influencers that we will demonstrate in the computational results is all the more
surprising.

\subsection{Parameterization and Model Creation.} The parameters in our model are $n$, $\lambda$, $\gamma$, $\delta$, $\alpha$, $b$, $\epsilon$,
$p_L$, and $p_R$.  Using the ODyN library, the {\tt OpionionNetworkModel} class can be initialized with these parameters as arguments.  This model can be populated with individuals bearing both weight and belief scores as described in the previous sections using {\tt populate\_model()}.  The belief propagation simulator is loaded as a separate class, {\tt NetworkSimulator,} and network simulations can be carried out on the model with {\tt run\_simulation()}.  Further documentation and demonstrations of this workflow can be found on the project Github page, and pseudocode for these procedures are given in Algorithms \ref{alg:populate} and \ref{alg:hk} below.

\section{Data Overview}\label{data_overview}
\subsection{CDC Vaccine Rate Data}

The Centers for Disease Control (CDC) vaccine trends data set shows overall trends in vaccination at the county level \cite{cdc_county_trends}. This data set includes counts and per-capita rates of first dose administration and full vaccination. For more comprehensive information about  the data reporting see  \cite{cdc_data}. The most current version of this data set can be loaded into ODyN using the function {\tt load\_county\_trend\_data} with the argument {\tt download\_data = True}.  This will pull the most up-to-date county vaccination data directly from the CDC website.

\subsection{ASPE Vaccine Hesitancy Data} 

The Assistant Secretary for Planning and Evaluation (ASPE) vaccine hesitancy data set gives county level estimates for Covid-19 vaccine hesitancy \cite{cdc_hesitancy_trends}.  State level estimates of vaccine hesitancy from the 2019 Household Pulse Survey (HPS) are combined with the Census Bureau's 2019 American Community Survey (ACS) 1-year Public Use Microdata Sample (PUMS) to obtain county level estimates of Covid-19 vaccine hesitancy. The HPS includes the survey question: {\em ``Once a vaccine to prevent Covid-19 is available to you, would you...get a vaccine?"}, with the responses 1) ``definitely get a vaccine", 2) ``probably get a vaccine", 3) ``unsure", 4) ``probably not get a vaccine" and 5) ``definitely not get a vaccine."  In the results of this survey, responses are reduced to a 3-point scale, classified as,
\begin{itemize}
\item {\em Strongly hesitant} if ``definitely not."
\item {\em Hesitant or unsure} if ``unsure" or ``probably not" or ``definitely not."
\end{itemize}
We assign a belief score of 2 to the ``strongly hesitant," 
a belief score of 1 to those who are ``hesitant or unsure" but not ``strongly hesitant," and a belief score of 0 to all others.  In this work we adopt the language and 3-point scale classification, since the raw data (i.e. 5-point) responses are not made available.
We use HPS data from Week 31, i.e., May 26, 2021 to June 7, 2021 \cite{aspe_week_31}.  This data set can be loaded into ODyN using the function {\tt load\_county\_hesitancy\_data} with the argument {\tt download\_data = True}.  This will load the Week 31 hesitancy data set directory from the CDC website.  Earlier data sets documenting vaccine hesitancy exist and have been used to examine trends in hesitancy at the county level \cite{aspe_week_26}. However, the CDC survey methodology changed, so older data  are not comparable to Week 31 data.

\section{Experimental Methodology} \label{methodology}

The model described in Section \ref{model} is generated by Algorithms \ref{alg:populate} and \ref{alg:hk} below.  To populate the network, we generate uniformly distributed random points in a triangle $T$.  

\begin{algorithm}
\caption{}\label{alg:populate}
\begin{algorithmic}[1]
\Procedure {{\tt populate\_model}}{$n$,$T$, $\gamma$, $(p_0,p_1,p_2)$, $\lambda$,$\alpha$, $b$, $\epsilon$, $p_L$, $p_R$}
\State $t_1,t_2,t_3 \gets$ vertices of triangle $T$.
\State ${\tt Agent} \gets $ empty $n\times 2$ position array
\State ${\tt Weight} \gets $ empty $n\times 1$ weight array
\State ${\tt Neighbor} \gets$ $n\times n$ array of zeros.
\State ${\tt MegaInfluences} \gets $ empty $n\times 2$ array.
\For{$i \leq n$}
\State $x,y,w \gets$ sampled from $U(0,1)$
\If{$x+y >1$ is even}
    \State $x \gets 1-x$
    \State $y \gets 1-y$  
\EndIf
\State ${\tt Agent}[i]\gets x \cdot (t_2-t_1) + y \cdot (t_3 - t_1)$
\State ${\tt Weight}[i] \gets $sampled from $U(0,1)$
\EndFor
\State ${\tt Weight} \gets {\tt Weight} ^ {-\frac{1}{\gamma}}$
\State ${\tt Belief} \gets$ $n\times 1$ array sampled from $[0,1,2]$ with probabilities $(p_0,p_1,p_2)$, resp.
\For{$i\leq n$} \label{get_neighbor}
\For{$j\leq n$ with $i\neq j$}
\State $x \gets$ sampled from $U(0,1)$
\If{$x < p_{u_iu_j}$ computed using Eq~\eqref{eqn:prob_influence}}
\State ${\tt Neighbor}[i,j] \gets 1$ 
\EndIf
\EndFor
\EndFor \label{end_get_neighbor}
\State $L \gets$ set of agents with belief within $\epsilon$ of the left influencer.
\State $R \gets$ set of agents with belief within $\epsilon$ of the right influencer. 
\For{$i$ in randomly sampled subset of $L$ with size $p_L \cdot \mid L\mid$}
\State ${\tt MegaInfluencer}[i,0]\gets 1$
\EndFor
\For{$i$ in randomly sampled subset of $R$ with size $p_R \cdot \mid R\mid$}
\State ${\tt MegaInfluencer}[i,1]\gets 1$
\EndFor
\State \textbf{return} ${\tt Agent},{\tt Weight},{\tt Belief}, {\tt Neighbor}, {\tt MegaInfluencer}$ 
\EndProcedure
\end{algorithmic}
\end{algorithm}

\subsection{Exploration of parameter space}
We run several experiments varying initial belief distributions, as well as the reach of influence from the left and right. Each of the experiments has $n=1000$ agents/vertices and parameters: $\lambda = 1/10$ the diameter of $T$, $\delta = 8.5$, $\alpha = 1.6$, $b = 1.5$ and $\epsilon = 1.5$.  These parameters were explicitly chosen to achieve clustering coefficients and in-degrees that were realistic for real-life community interactions, namely, a consistent clustering coefficient of approximately 0.3 as well as an average in-degree between 7 and 9 (for populations that begin with a strong consensus belief, we expect marginally higher average degree). Though a person might interact with a larger number of individuals through their online social networks, or a smaller number of individuals through in-person interactions, surveys have shown that people report feeling genuinely close to between 5 and 10 individuals in their social circle, broadly construed \cite{dunbar}. The clustering coefficient was chosen to be consistent with values for average clustering coefficients on directed graphs using random walks on social networks \cite{hardiman}. As noted
earlier, when generating the weights $W_u$, we used $\gamma=1.5$ which yields a heavy-tailed distribution that has finite mean but infinite variance. Due to computational constraints, in the present paper we do not carry out a full exploration of parameter space, although this is an interesting next step in the research. But our limited selection of parameters were chosen to mimic a real-life network in a way that's quantitatively supported by social science research as mentioed above. Similary we restrict our attention to networks with only 1000  nodes, bearing in mind that such networks may be susceptible to edge effects.

\begin{algorithm}
\caption{}\label{alg:hk}
\begin{algorithmic}[1]
\Procedure {{\tt run\_simulation}}{$n$,$\Delta$, $\gamma$, $(p_0,p_1,p_2)$, $\lambda$,$\alpha$, $b$, $\epsilon$, $p_L$, $p_R$}
\State Compute ${\tt Agent},{\tt Weight},{\tt Belief}, {\tt Neighbor}, {\tt MegaInfluencer}$ with Algorithm \ref{alg:populate}.
\While{ {\tt True}}
\For{$j\leq n$}
\State $S\gets \{{\tt Belief}[j]\}\cup \{{\tt Belief}[k] :{\tt Neighbor}[k,j] =1 \text{ for }k\neq j\}$
\If{${\tt MegaInfluencer}[j,0] = 1$} \label{mega_inf} 
\If{${\tt Belief}[j] < \epsilon$}
\State \# Include left mega-influencer belief in $S$.
\State $S\gets S\cup \{\text{left mega-influencer belief (i.e. 0)}\}$
\Else 
\State \# Connect to right mega-influencer with probability $p_R$.
\State $x \gets$ sampled from $U(0,1)$
\If{$x< p_R$}
\State ${\tt MegaInfluencer}[j,0] \gets 0$ and ${\tt MegaInfluencer}[j,1] \gets 1$
\EndIf 
\EndIf
\EndIf \label{end_mega_inf}
\State Repeat steps \ref{mega_inf} - \ref{end_mega_inf} for right mega influencer.
\State \# Propagate all local beliefs. 
\State ${\tt Belief}[j] \gets$ average of beliefs in $S$.
\EndFor
\State \# Recompute network graph using updated beliefs.
\State ${\tt Neighbor} \gets$ Recompute ${\tt Neighbor}$ using steps \ref{get_neighbor} - \ref{end_get_neighbor} in Algorithm \ref{alg:populate}.
\If{{\tt stopping\_criterion\_satisfied}}
\State {\tt break}
\EndIf
\EndWhile
\State \textbf{return} ${\tt Belief}$
\EndProcedure
\end{algorithmic}
\end{algorithm}

As our initial belief distributions, we consider four distinct cases as shown in Table \ref{table:initial_beliefs}.  For the {\em Vaccine Accepting} population we use the initial hesitancy beliefs of Multnomah County, Oregon, and we negate these beliefs to seed a {\em Vaccine Skeptical} population.  The {\em Skewed} beliefs are distributed according to the vaccine hesitancy beliefs in Montgomery County, Alabama, which skew towards acceptance, but with a significant skeptical population.  Finally, we include a fictitious population with {\em Symmetric} beliefs to demonstrate the outcomes in a symmetric population. Using these distributions, we seed our model using Algorithm \ref{alg:populate} to obtain the initial network models shown in Fig~\ref{fig:initial_seeded_models}.

With Algorithm \ref{alg:populate} we assign attributes to individual agents such as weight, spatial distance, position in opinion space, and connection to mega-influencers, which we then use to compute the network graph.  Using Algorithm \ref{alg:hk} we synchronously update all opinions.  After each round of opinion updates, the network graph is recomputed holding weight and spatial distance parameters constant.

\subsection{Stopping criterion}

\begin{figure}
\includegraphics[width = .8\linewidth]{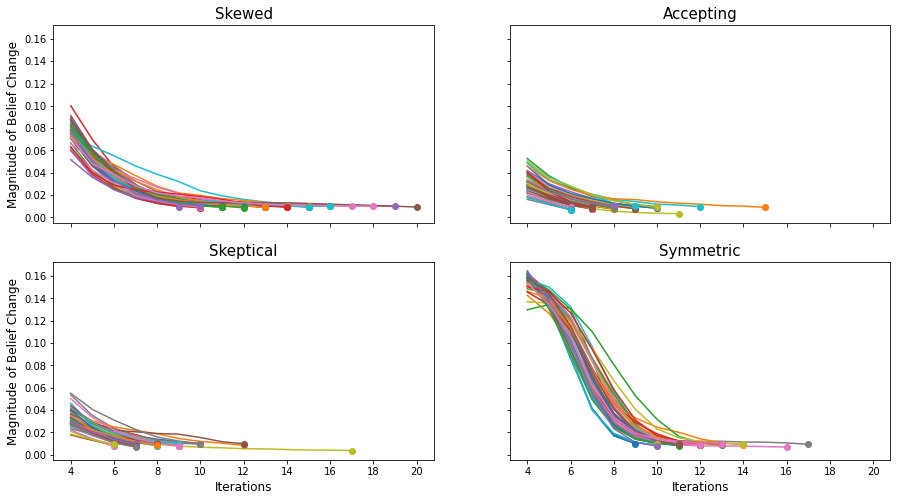}
\caption{Mean change in absolute belief over timesteps.  The simulation is stopped when the community-wide mean of absolute change in belief drops below a .01 across a 5-time step rolling average.}
\label{fig:stopping}
\end{figure}

For each initial belief distribution, we carry out seven experiments, using Algorithm \ref{alg:hk}, varying the left and right mega-influence (i.e., varying $p_L$ and $p_R$).  In one experiment we set both of $p_L$ and $p_R$ equal to 0 to simulate a population without any influence from mega-influencers.  In the remaining six experiments, $p_R = 0.8$ and $p_L = 0,.2,.4, .6, .8$ or $1$. A stopping criterion is determined as follows.  At each timestep, a 5-time step rolling average in belief change is calculated for each individual.  The community-wide mean of the absolute change in belief is then computed.  When this value drops below .01, the simulation is stopped.  We note that this allows for individuals to have small oscilations in opinion, but overall the community opinion stabilizes.  For brevity, in Algorithm \ref{alg:hk} we indicate this with a boolean {\tt stopping\_criterion\_satisfied}. In Fig \ref{fig:stopping}, we can see that this threshold is typically reached in 20 or fewer time steps.

\begin{figure}
\centering
  \includegraphics[width=5in]{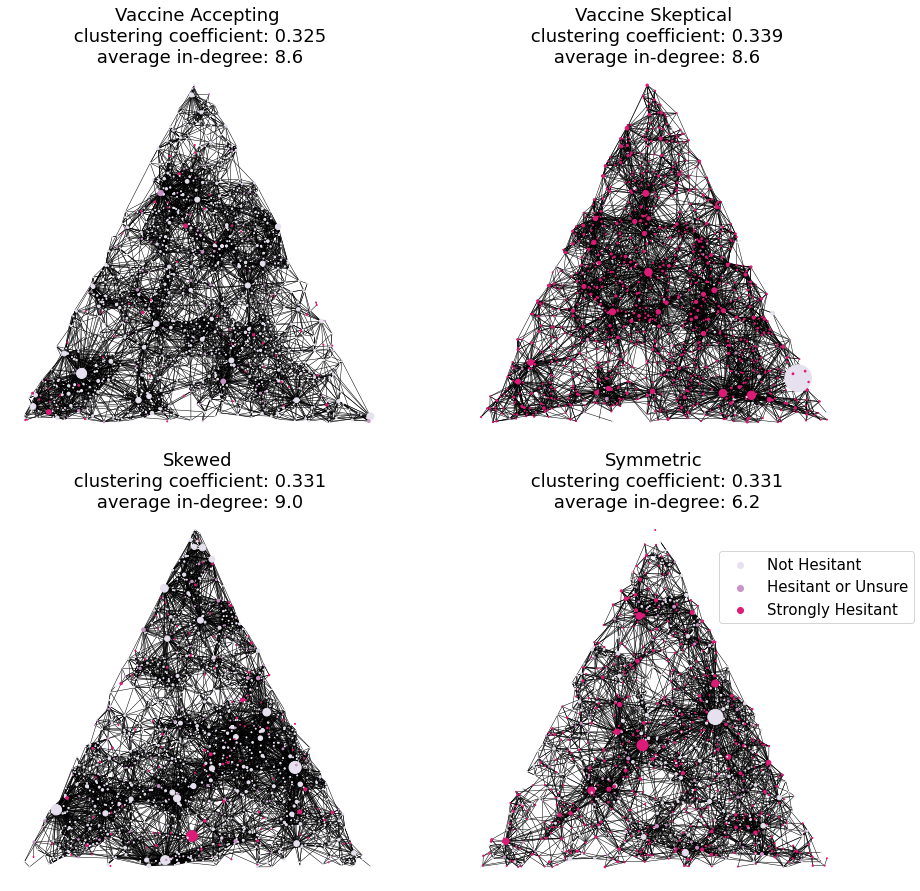}
  \caption{Initialized networks with varying initial belief distributions from Table \ref{table:initial_beliefs}, for $n=1000$ individuals and connections based on weight and distance in both physical and opinion space.}
  \label{fig:initial_seeded_models}
\end{figure}

After each time step the network adjacencies are re-computed according to the updated beliefs. This reflects the
fact that somebody who influences my opinion today may not influence it tomorrow, for instance because I may happen not to talk to them tomorrow.
Fig~\ref{fig:initial_seeded_models} depicts examples of the initialized networks for each of the four scenarios we analyzed. For each of the scenarios the average-in degree was between 6 and 9, which corresponds to realistic social networks \cite{dunbar,hardiman}. 
Our simulation results are presented in Section \ref{results}. 

\begin{center}
\begin{table}
\begin{tabular}{cccccc}
    Belief Type &  $p_0$ & $p_1$ & $p_2$ & Mean Belief &Corresponding US County\\
     \hline
     Vaccine Accepting & $0.8831$ & $0.0492$ & $0.0677$ & $0.1846$ &Multnomah County, OR\\
     Vaccine Skeptical & $0.0677$ & $0.0492$ & $0.8831$ & $1.8154$ &\\
    Skewed & $0.7877$ &  $0.0988$ & $0.1135$ & $0.3258$ &Montgomery County, AL\\
     Symmetric &  $ 0.4500$ & $0.1000$ & $0.4500$ & $1.0000$&\\
     \hline 
\end{tabular}
\caption{Initial Population Belief Distributions}\label{table:initial_beliefs}
\end{table}
\end{center}

\section{Results}\label{results}

\begin{figure}
\includegraphics[width = .8\linewidth]{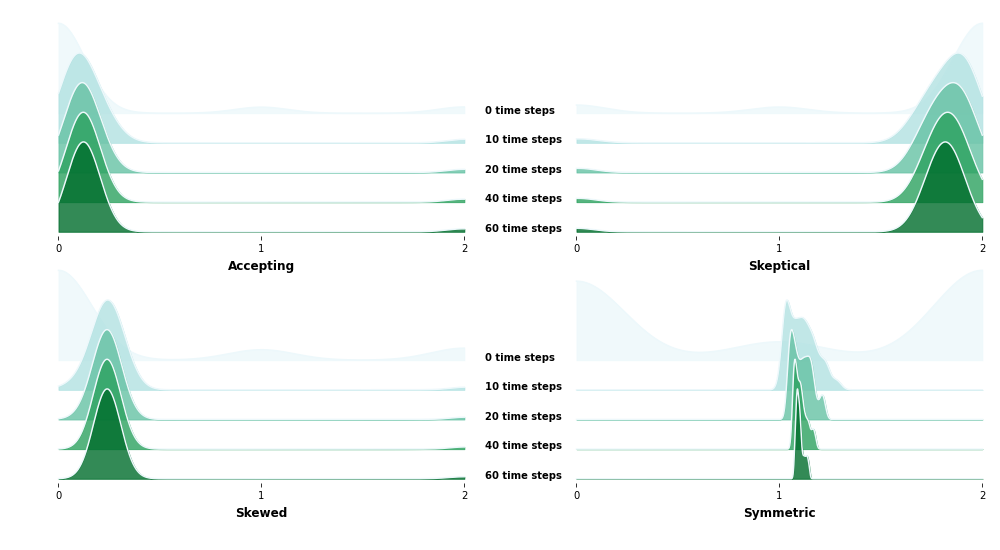}
\caption{Populations with initial belief distributions from Table \ref{table:initial_beliefs} over 60 time steps without mega-influencers.}
\label{fig:ridge_none}
\end{figure}

\begin{figure}
    \centering
    \includegraphics[width = 4in]{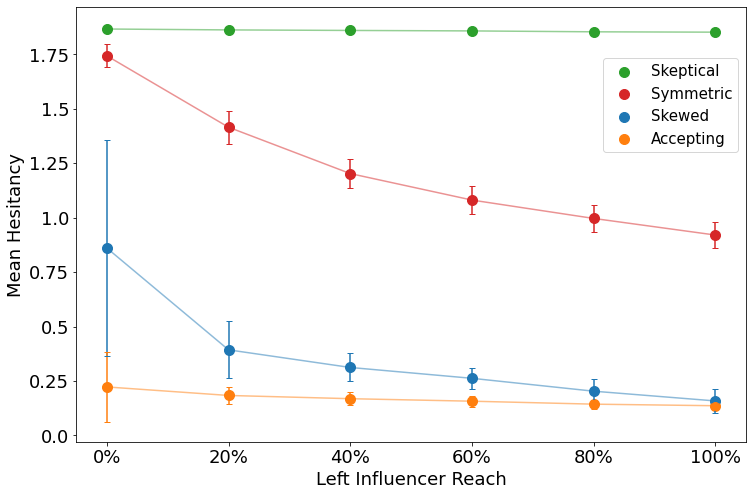}
    \caption{Mean belief at stopping as a function of left influencer reach for four different initial conditions.  Averaged over 50 runs with errors bars indicating 1 standard deviation. The {\em Heavily Skewed Accepting} and {\em Skeptical} populations begin with initial vaccine hesitancy estimates of Multnomah County, OR and its reflection.  The {\em Slightly Skewed Accepting} population begins with initial vaccine hesitancy estimates of Montgomery County, AL.}
    \label{fig:phase_shifts}
\end{figure}

We visualize our simulations using smoothed ridge plots with a Gaussian kernel density estimate.  Results  with $p_L=0$ and $p_R=0$ (no mega-influencers) are shown in Fig~\ref{fig:ridge_none}.  We see that 
without mega-influencers, populations tend to reach tight  consensus near the mean of the
initial opinion distribution.  For clarity and ease of comparison, in this section we show the results for one simulation that was carried for precisely 60 timesteps (well beyond the stopping criterion for all scenarios); these artificially truncated simulations still provide statistically sound examples of the the overall trends from all simulations, which is further confirmed by the aggregate statistics presented in Fig~\ref{fig:phase_shifts}.
In Hegselmann-Krause dynamics, often multiple clusters are seen, not just 
a single peak \cite{hegsel_krause_2002}. In our simulations, we invariably see 
a single peak forming. This is because we use a fairly large value of the parameter 
$b$; that is, in our model, people are fairly willing to listen to others with somewhat 
diverging views, and be influenced by those views.

\begin{figure}
\includegraphics[width = .8\linewidth]{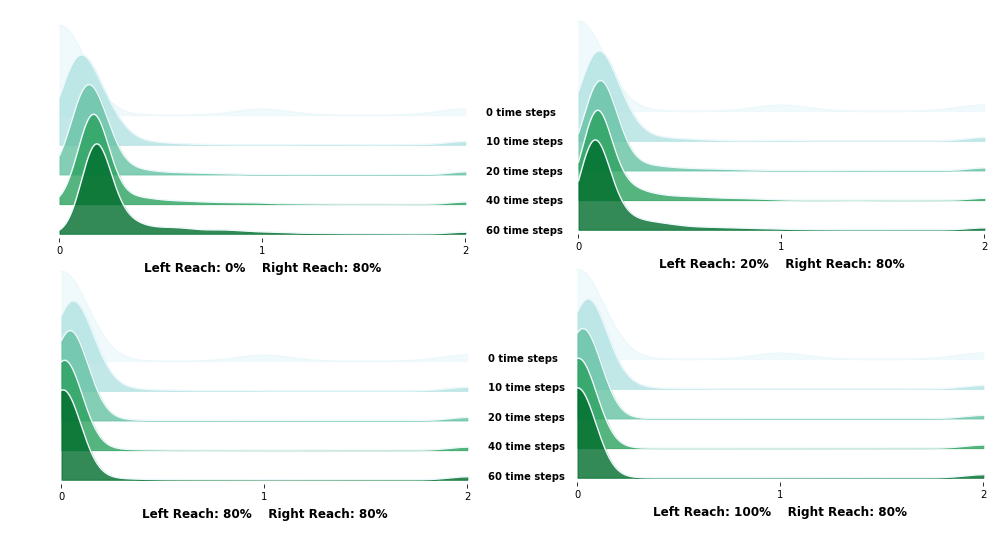}
\caption{A population with {\em accepting} initial beliefs (such as Multnomah County, Oregon) over 60 time steps with a right mega-influencer of strong reach ($p_R=0.8$) and a left mega-influencer
of varying reach ($p_L=0,0.2,0.8$,and $1$).}
\label{fig:or_ridge}
\end{figure}

\begin{figure}
\includegraphics[width = .8\linewidth]{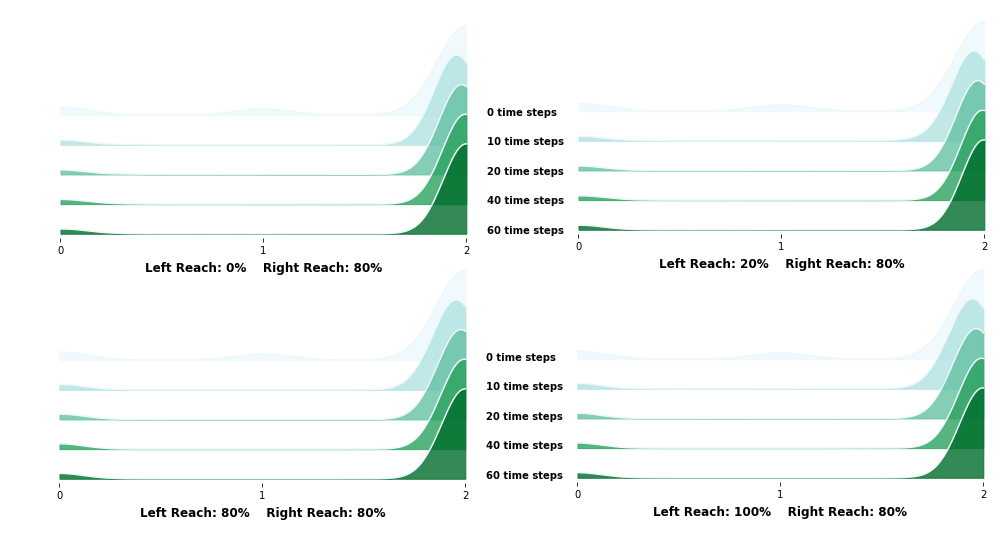}
\caption{A population with {\em skeptical} initial beliefs over 60 time steps with a right mega-influencer of strong reach ($p_R=0.8$) and a left mega-influencer
of varying reach ($p_L=0,0.2,0.8$,and $1$).}
\label{fig:anti_ridge}
\end{figure}

\begin{figure}
\includegraphics[width = .8\linewidth]{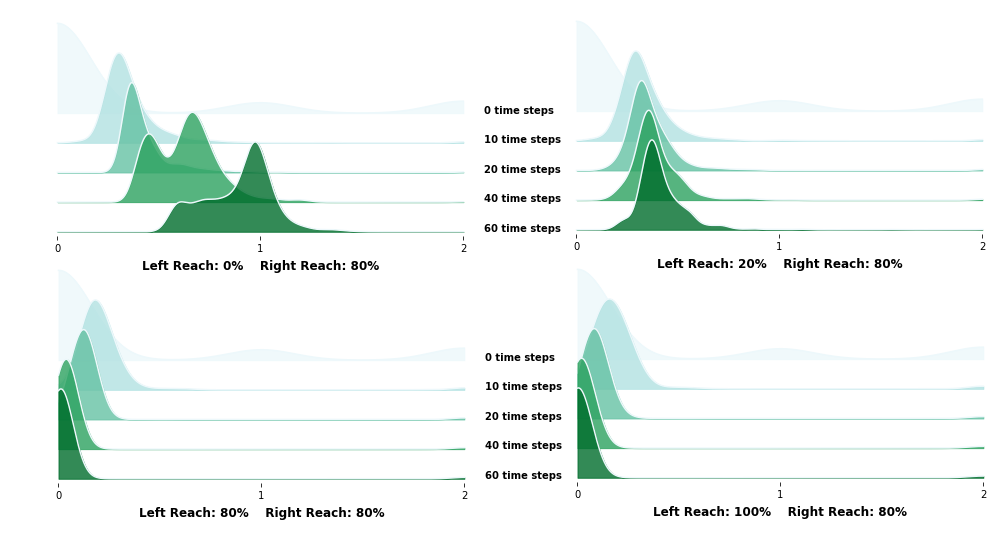}
\caption{A population with {\em skewed} initial beliefs (such as Montgomery County, Alabama) over 60 time steps with a right mega-influencer of strong reach ($p_R=0.8$) and a left mega-influencer
of varying reach ($p_L=0,0.2,0.8$,and $1$).}
\label{fig:al_ridge}
\end{figure}

\begin{figure}
\includegraphics[width = .8\linewidth]{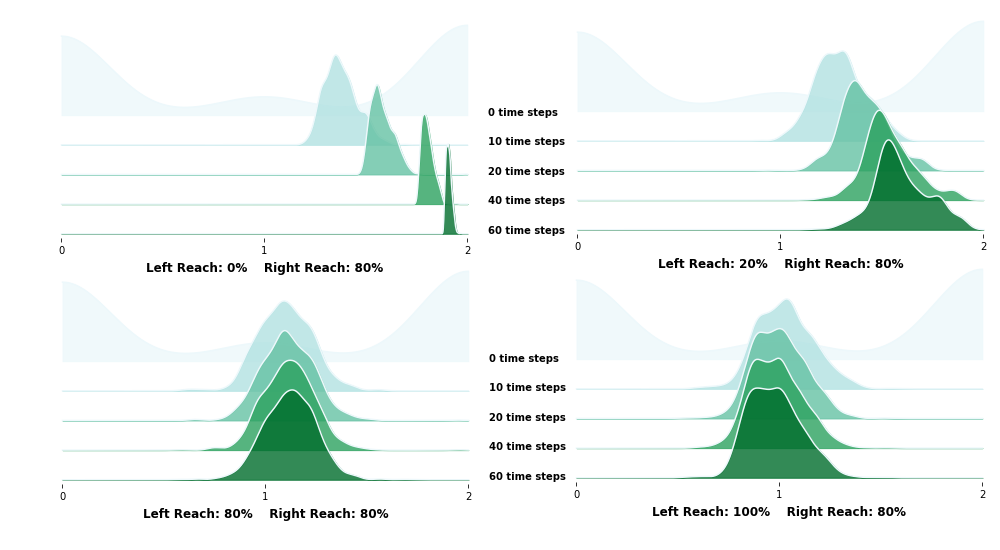}
\caption{A population with {\em symmetric} initial beliefs over 60 time steps with a right mega-influencer of strong reach ($p_R=0.8$) and a left mega-influencer
of varying reach ($p_L=0,0.2,0.8$,and $1$).}
\label{fig:sym_ridge}
\end{figure}

Now we set $p_R$ (the reach of the right influencer) to 80$\%$ and vary the reach of the left influencer to see how this impacts the long term opinions of variously seeded populations.  These results are shown for the Vaccine Accepting (Fig~\ref{fig:or_ridge}), Vaccine Skeptical (Fig~\ref{fig:anti_ridge}), Skewed (Fig~\ref{fig:al_ridge}), and Symmetric (Fig~\ref{fig:sym_ridge}).  

In the Vaccine Accepting population, we see that even in the absence of mega-influence from the left, the population remains relatively vaccine accepting over 60 time steps although there is some diffusion of opinion.  With a modest increase of $20\%$ reach by the left mega-influencers, vaccine acceptance remains the dominating consensus belief. Unsurprisingly, the equivalent but opposite behavior is witnessed in the Vaccine Skeptical population.  Of course there is slight variance in the two outcomes due to the stochastic nature of the simulation.   

This behavior is summarized in Fig~\ref{fig:phase_shifts}, which shows an average across 50 runs
 of the mean belief at stopping, where the left influencer's reach is changing but the right one is kept at $80\%$ reach. In a vaccine skeptical population, given sufficient mega-influence from the right, no amount of left influence is capable of changing vaccine beliefs in any meaningful way.  However, critically, we see that in skewed populations (such as we see in Mongtomery County, AL) a change in left influence has a dramatic impact on the overall vaccine acceptance rate.  With a left influencer reach of $60\%$ the mean belief is around $0.25$       which suggests a willingness to receive a vaccine.

The experiments can be easily replicated in the command line with scripts included in ODyN: {\tt accepting\_simulation.py}, {\tt skeptical\_simulation.py}, {\tt symmetric\_simulation.py} and {\tt skewed\_simulation.py}.  The data corresponding to these results is stored on Github at \url{https://github.com/annahaensch/ODyN/tree/main/data/simulations_from_paper} and the ODyN repository also includes a Jupyter notebook to recreate the plots shown in this paper.

\section{Conclusion} \label{conclusion}

The most interesting scenario among those we have simulated is that of a population
that initially leans towards accepting the vaccine, with some vaccine skeptics. 
The data suggest that this may have been the situation in the state of Alabama,
for instance, when Covid-19 vaccines were first developed. 

Remarkably, in such a situation, if there were no mega-influencers, tight vaccine-accepting consensus would emerge in our model. However, a vaccine-hesitant mega-influencer who 
reaches a large 
fraction of those who are not strongly vaccine-accepting to begin with can cause
a dramatic shift towards vaccine hesitancy. 
This is true even under our very conservative assumption that
a mega-influencer counts, for those who listen to them, no more than a single
friend, colleague, or neighbor would count. 

A competing vaccine-accepting mega-influencer reaching all those 
who are not strongly vaccine-hesitant already can 
counter-act this effect. In fact, it is surprisingly easy for the vaccine-accepting
mega-influencer to reverse the trend: They do not need to match the reach of the
vaccine-hesitant mega-influencer to succeed. 

However, while the vaccine-accepting mega-influencer can shift the mean population view
back towards vaccine acceptance, a much greater spread in views is always
seen 
when there are two mega-influencers, in comparison with a population in which 
there are none. 

We summarize our main points. 

\begin{itemize}
\item[-] Powerful mega influence from the right can move even a strongly-vaccine {\em accepting} population over to the right. Just talking to the people around you (friends, family, etc.) is not enough to effect widespread community belief change.  
\item[-] You can stop mass-migration to the right, but you may not be able to stop broadening of the belief distribution, and this is especially noticeable in initially {\em skewed} populations.
\end{itemize}

Social media make spatial proximity less important,  but tend to 
make people interact more selectively with the like-minded as both a consequence of social and algorithmic behavioral drivers \cite{cinelli_social}. 
One could attempt to model this effect by changing parameters in our model, making
spatial proximity less important, and making like-mindedness more important, 
in determining the probabilities $p_{uv}$.

Our results also suggest future work on the dependence on the parameters 
$b$ and $\epsilon$; note that smaller values of $b$ and $\epsilon$ will require
replacing the discrete initial distributions (beliefs of 0, 1, or 2) by a continuous
distribution. Further, we intend to work on scalable sampling algorithms for combining triangles into counties
and counties into
states. We plan to attempt to understand how time steps in our model map onto real 
time. Another feature to be added to the model in the future would be the effects of 
central interventions such as government or workplace vaccine mandates. The connection between beliefs and Covid-19 is also seen in the study of the spread of misinformation \cite{rabb_plosone}; these notions could also be incorporated into our model. 

\section{Acknowledgments}
The authors wish to thank the Data Intensive Studies Center at Tufts University for generous support through a SEED Grant.

\bibliographystyle{plain}

\begin{thebibliography}{1}

\bibitem{review_paper}
A. Sirbu, V. Loreto, V. D. P. Servedio and F. Tria. 
\newblock \textit{Opinion dynamics: models, extensions and external effects}.  
\newblock 2016, \url{https://arxiv.org/abs/1605.06326}.

\bibitem{hofstad_networks_book}
R. van der Hofstad. 
\newblock \textit{Random Graphs and Complex Networks, Volume 1}.
\newblock Cambridge Series in Statistical and Probabilistic Mathematics, 2017.

\bibitem{survey_general}
J. Lorenz. 
\newblock \textit{Continuous opinion dynamics under bounded confidence: A survey}. 
\newblock Int. J. Modern Phys. C, vol. 18, no. 12, pp. 1819–1838, 2007.

\bibitem{cdc_hesitancy_trends}
Centers for Disease Control and Prevention.
\newblock \textit{Vaccine Hesitancy for COVID-19: County and local estimates}.
\newblock \url{https://data.cdc.gov/Vaccinations/Vaccine-Hesitancy-for-COVID-19-County-and-local-es/q9mh-h2tw}, Last Accessed: December 8, 2021.

\bibitem{lengyel_2015}
Lengyel, B., Varga, A., Ságvári, B., Jakobi, Á., and Kertész, J.
\newblock {\em Geographies of an online social network}. 
\newblock PloS one 10.9 (2015): e0137248.

\bibitem{cdc_county_trends}
Centers for Disease Control and Prevention.
\newblock \textit{COVID-19 Vaccinations in the United States, County}.
\newblock \url{https://data.cdc.gov/Vaccinations/COVID-19-Vaccinations-in-the-United-States-County/8xkx-amqh}, Last Accessed: December 8, 2021.

\bibitem{cdc_national_trends}
Centers for Disease Control and Prevention. 
\newblock \textit{Trends in Number of COVID-19 Vaccinations in the US}.
\newblock\url{https://covid.cdc.gov/covid-data-tracker/#vaccination-trends}, Last Accessed: December 8, 2021.

\bibitem{cdc_data} 
Centers for Disease Control and Prevention. 
\newblock \textit{COVID-19 Vaccination}. 
\newblock \url{https://www.cdc.gov/coronavirus/2019-ncov/vaccines/distributing/about-vaccine-data.html}, Last Accessed: December 8, 2021.

\bibitem{murthy_rural} 
B. P. Murthy, N. Sterrett, D. Weller, et al. 
\newblock \textit{Disparities in COVID-19 Vaccination Coverage Between Urban and Rural Counties -- United States}.
\newblock December 14, 2020–April 10, 2021. MMWR Morb Mortal Wkly Rep 2021;70, pp. 759-- 764. DOI: \url{http://dx.doi.org/10.15585/mmwr.mm7020e3external icon}.

\bibitem{simas_overcoming}
C. Simas and H. J. Larson. 
\newblock \textit{Overcoming vaccine hesitancy in low-income and middle-income regions}. 
\newblock Nat Rev Dis Primers 7, 41 (2021). 
\url{https://doi.org/10.1038/s41572-021-00279-w}

\bibitem{rabb_plosone}
N. Rabb, L. Cowen, J. P. de Ruiter, M. Scheutz, M. 
\newblock {\em Cognitive cascades: How to model (and potentially counter) the spread of fake news. }
\newblock Plos one, 17(1), e0261811. 

\bibitem{beatty_2021}
A.L.Beatty, N.D. Peyser, X.E. Butcher, et al. 
\newblock {\em Analysis of COVID-19 Vaccine Type and Adverse Effects Following Vaccination}. 
\newblock JAMA Netw Open. 2021;4(12):e2140364. 
\newblock doi:10.1001/jamanetworkopen.2021.40364

\bibitem{kriss_completion}
J. L. Kriss, L. E. Reynolds, A. Wang, et al. 
\newblock \textit{COVID-19 Vaccine Second-Dose Completion and Interval Between First and Second Doses Among Vaccinated Persons -- United States}. 
\newblock December 14, 2020 -- February 14, 2021. MMWR Morb Mortal Wkly Rep 2021;70:389–395. DOI: \url{http://dx.doi.org/10.15585/mmwr.mm7011e2}.

\bibitem{hegsel_krause_2002}
R. Hegselmann and U. Krause.
\newblock \textit{Opinion dynamics and bounded confidence models, analysis, and simulation}.
\newblock Journal of Artificial Societies and Social Simulation, vol. 5, 2002.

\bibitem{hegsel_krause_2015}
R. Hegselmann and U. Krause.
\newblock {\em Opinion dynamics under the influence of radical groups, charismatic leaders, and other constant signals: A simple unifying model}. 
\newblock Networks \& Heterogeneous Media 10.3 (2015): 477.


\bibitem{chung_lu1}
F. Chung and L. Lu.
\newblock \textit{The average distances in random graphs with given expected degrees}. 
\newblock Proceedings of the National Academy of Sciences (PNAS), 99(25), pp. 15879 -- 15882, 2002.

\bibitem{chung_lu2}
F. Chung and L. Lu.
 \newblock \textit{Connected components in random graphs with given expected degree sequences}. 
 \newblock Annals of Combinatorics, 6(2), pp. 125 -- 145, 2002.

\bibitem{chung_lu3}
F. Chung and L. Lu.
 \newblock \textit{The average distance in a random graph with given expected degrees}. 
 \newblock Internet Mathematics, 1(1), pp. 91 -- 113, 2004.

\bibitem{girgs_algorithm_paper}
K. Bringmann, R. Keusch, and J. Lengler. 
\newblock \textit{Sampling Geometric Inhomogeneous Random Graphs in Linear Time}.  
\newblock 2016, \url{https://arxiv.org/abs/1511.00576}.

\bibitem{mathias_2016}
J-D. Mathias, S. Huet, and G. Deffuant. 
\newblock {\em Bounded confidence model with fixed uncertainties and extremists: The opinions can keep fluctuating indefinitely}. 
\newblock Journal of Artificial Societies and Social Simulation 19.1 (2016): 6.

\bibitem{barthelemy_2011}
M. Barthélemy. 
\newblock {\em Spatial networks}. 
\newblock Physics reports 499.1-3 (2011): 1-101.

\bibitem{literature_hk1}
U. Krause.
\newblock \textit{Soziale Dynamiken mit vielen Interakteuren. Eine Problemskizze, in Proc. Modellierung Simul. von Dynamiken mit vielen interagierenden Akteuren}. 
\newblock 1997, pp. 37–1.

\bibitem{literature_hk2}
U. Krause.
\newblock \textit{A discrete nonlinear and non-autonomous model of consensus formation}. 
\newblock Communications in difference equations, 2000 (2000), pp. 227 -- 236.

\bibitem{literature_hk3}
J. Lorenz.
\newblock \textit{A stabilization theorem for continuous opinion dynamics}. 
\newblock Physica A, vol. 355, no. 1, pp. 217–223, 2005.

\bibitem{literature_hk4}
J. Lorenz.
\newblock \textit{Consensus strikes back in the Hegselmann-Krause model of continuous opinion dynamics under bounded confidence}.
\newblock J. Artif. Societies Social Simul., vol. 9, no. 1, 2006.

\bibitem{other_model1}
E. Ben-Naim.
\newblock \textit{Rise and fall of political parties}.
\newblock Europhys. Lett., vol. 69, no. 5, pp. 671–676, 2005.

\bibitem{other_model2}
S. Fortunato, V. Latora, A. Pluchino and A. Rapisarda. 
\newblock \textit{Vector opinion dynamics in a bounded confidence consesnus model}. 
\newblock Int. J. Modern Phys. C, vol. 16, pp. 1535–1551, Oct. 2005.

\bibitem{other_model3}
D. Urbig. 
\newblock \textit{Attitude dynamics with limited verbalisation capabilities}. 
\newblock J. Artif. Societies Social Simul., vol. 6, no. 1, 2003.

\bibitem{gargiulo_2016}
F. Gargiulo and Y. Gandica. 
\newblock {\em The role of homophily in the emergence of opinion controversies}. 
\newblock arXiv preprint (2016) arXiv:1612.05483.



\bibitem{aspe_week_31} 
Department of Health and Human Services.
\newblock \textit{ASPE Predictions of Vaccine Hesitancy for COVID-19 Vaccines by Geographic and Sociodemographic Features}. 
\newblock Methodological Description, Assistant Secretary for Planning and Evaluation Issue Brief, June 17, 2021, 
\url{https://aspe.hhs.gov/sites/default/files/migrated_legacy_files//200821/vaccine-hesitancy-COVID-19-Methodology.pdf},
Last Accessed: Oct. 20, 2021.

\bibitem{aspe_week_26} 
T. Beleche, J. Ruhter A. Kolbe, J. Marus, L. Bush and B. Sommers. 
\newblock \textit{COVID-19 Vaccine Hesitancy: Demographic Factors, Geographic Patterns, and Changes Over Time}.
\newblock Assistant Secretary for Planning and Evaluation Issue Brief, Department of Health and Human Services, 
May, 2021, \url{https://aspe.hhs.gov/sites/default/files/migrated_legacy_files//200816/aspe-ib-vaccine-hesitancy.pdf}, 
Last Accessed: Oct. 13, 2021.

\bibitem{dunbar}
R.I.M. Dunbar. 
\newblock \textit{Do online social media cut through the constraints that limit the size of offline social networks?}. 
\newblock Royal Society Open Science 3.1 (2016): 150292.

\bibitem{hardiman}
S. J. Hardima and L. Katzir. 
\newblock \textit{Estimating clustering coefficients and size of social networks via random walk}. \newblock Proceedings of the 22nd international conference on World Wide Web. 2013.

\bibitem{cinelli_social}
M. Cinelli, G. De Francisci Morales, A. Galeazzi, W. Quattrociocchi and M. Starnini. 
\newblock \textit{The echo chamber effect on social media}.
\newblock Proceedings of the National Academy of Sciences Mar 2021, 118 (9) e2023301118; DOI: 10.1073/pnas.2023301118


\bibitem{social_media}
M. Workman. 
\newblock \textit{An empirical study of social media exchanges about a controversial topic: Confirmation bias and participant characteristics}.  
\newblock Journal of Media in Society, 7 (1), 2018, pp. 381--400.


\end{thebibliography}

\end{document}